# Carrier dynamics in silicon nanowires studied via femtosecond transient optical spectroscopy from 1.1 to 3.5 eV.

Lin Tian[1,#][田琳], Lorenzo Di Mario[1,2], Aswathi K. Sivan[1], Daniele Catone[2], Patrick O'Keeffe[3], Alessandra Paladini[3], Stefano Turchini[2] and Faustino Martelli[1]

[1]Istituto per la Microelettronica e i Microsistemi (IMM), CNR, 00133 Rome, Italy

[2] Istituto di Struttura della Materia-CNR (ISM-CNR), Division of Ultrafast Processes in Materials (FLASHit), Via del Fosso del Cavaliere, 100, 00133 Rome, Italy

[3] Istituto di Struttura della Materia-CNR (ISM-CNR), Division of Ultrafast Processes in Materials (FLASHit), 00016 Monterotondo Scalo, Italy

e-mail: faustino.martelli@cnr.it
# Present address: Institute for Quantum Computing and Department of Electrical & Computer Engineering, University of Waterloo, ON N2L 3G1, Waterloo, Canada

**Abstract** We present femtosecond transient transmission (or absorbance) measurements in silicon nanowires in the energy range 1.1-3.5 eV, from below the indirect band-gap to above the direct band-gap. Our pump-probe measurements allow us to give a complete picture of the carrier dynamics in silicon. In this way we perform an experimental study with a spectral completeness that lacks in the whole literature on carrier dynamics in silicon. A particular emphasis is given to the dynamics of the transient absorbance at the energies relative to the direct band gap at 3.3 eV. Indeed, the use of pump energies below and above 3.3 eV allowed us to disentangle the dynamics of electrons and holes in their respective bands. The band gap renormalization of the direct band gap is also investigated for different pump energies. A critical discussion is given on the results below 3.3 eV where phonon-assisted processes are required in the optical transitions.

KEYWORDS Si nanowires, ultrafast spectroscopy, carrier dynamics, band structure

**1. Introduction**

Despite silicon being the most important material for present-day technology and despite the extensive literature that deals with its optical properties, several aspects related to the carrier dynamics have not been fully characterized. This is mainly due to the fact that silicon has an indirect fundamental band gap at 1.1 eV that makes the related optical transitions weak and the scarcity of experimental systems that allow for time-resolved measurements over the full energy range covering the gap between indirect and direct band-edges [1].

As mentioned, most of the works that explore the carrier dynamics in Si have been performed via reflectivity [1-5]. The early works were focused on the effect that high densities (>$10^{21}$ cm$^{-3}$) of excited carriers induce on lattice dynamics, like melting or phase transitions [2,3], while later works more often deal with the non-equilibrium optical properties and carrier-phonon interaction [1, 4-8]. Most of the works present measurements in the visible region, with some of them addressing the near UV region with results on the direct band gap at 3.3 eV at room temperature. Some of the cited works report results on the band gap renormalization, due to the many-body effects occurring at high carrier densities. In particular, Schultze et al. [7] measured the energy redshift of the 2p core transition with attosecond resolution and assigned it as being due to the shrinking of the indirect Si band gap following carrier excitation by an intense few-cycle laser pump pulse. The renormalization of the 3.3 eV direct band gap was investigated in bulk material by Sangalli and coworkers [5], who compare theoretical calculations with experimental reflectivity results obtained at a single pump-probe delay (200 fs), when the reflectivity variation showed its maximum. More experimental detail is given in a further work of the same group [9].

Recent studies on Si nanowires (NWs) have mainly focused on the dependence of carrier dynamics on the NW diameter and in particular on the surface recombination velocity [8, 10-13]. The spectral region of the study was restricted to the visible region, 400-800 nm, not allowing for the study of the energies related to the indirect and direct band gaps and of the contribution of band gap renormalization to the transient spectra.

In this work, we report on femtosecond transient absorbance spectroscopy (FTAS) on Si, specifically on Si NWs, performing pump-probe transmission measurements over the wide spectral region 1.1-3.5 eV for the probe. The measurements have allowed us to reveal features previously unknown about the carrier dynamics in Si and its dependence on the band structure of silicon. Among other results we report on the gap renormalization dynamics of the direct gap and the disentanglement of the electron dynamics in the $\Gamma_{15}$ conduction band from the hole dynamics in the valence band, obtained by comparing the transient absorbance (TA) of the probe signal using pump energies above or below the direct band gap energy. The use of Si in the form of nanowires allows simple TA measurements on as-grown samples, without the need of using difficult preparation techniques of very thin layers starting from bulk material. The NW diameter (about 80 nm) is such that quantum confinement is not present and the results are hence expected to reflect the properties of bulk silicon. Our work is hence not only a work on the specific properties of Si NWs, but also a work that gives a more complete picture of carrier dynamics in silicon.

## 2. Experimental

Indium-seeded NWs have been grown by plasma-enhanced chemical vapor deposition at 400 °C on quartz. The growth results in NWs of high crystallinity [14,15]. The as-grown samples show high transparency as shown in Figure 1(a). The Si NWs are only grown in the round pattern, where the metal seed was deposited. Due to light trapping in the NWs array [16], the pattern appears darker than the surrounding region, covered by amorphous materials in a similar quantity. Scanning electron microscopy (Figure 1(b)) reveals a high density of randomly oriented, straight Si NWs, 1 μm long and with a diameter of about 80 nm. Au-seeded NWs, grown under otherwise similar conditions, have been also used for this study. As the results do not differ in any significant way from those obtained with In-seeded NWs, for sake of simplicity we only report the results obtained with the latter.

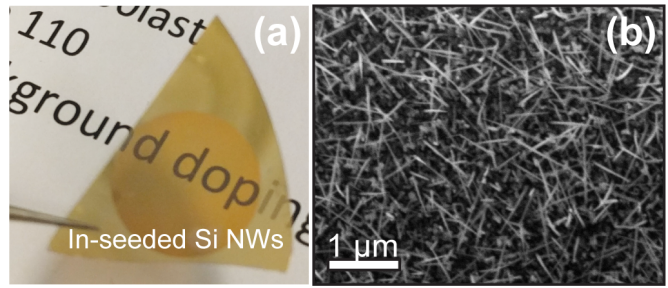

**Figure 1**. (a) Photo of In-seeded Si NWs showing high transparency, the round pattern is covered with Si NWs. (b) SEM image of In-seeded Si NWs.

FTAS was performed by pump-probe experiments at room temperature on as grown samples. As the pump, we used the second harmonic of an amplified Ti:sapphire laser at 405 nm (3.06 eV) or the 275 nm (4.51 eV) and 530 nm (2.34 eV) wavelengths produced by an optical parametric amplifier. As probe we used the supercontinuum white light generated in a femtosecond transient absorption spectrometer of IB Photonics (FemtoFrame II). The pump-probe response was studied in the 350–1600 nm wavelength range using two different crystals to generate the supercontinuum in the 800-1600 nm (a static YAG crystal) and 350-800 nm (a rotating $CaF_2$ crystal) ranges. With the pump energies ($E_p$) of 3.06 eV and 4.51 eV we used pump fluencies of 150 and 300 μJ/cm², while for $E_p$=2.34 eV, because of the low absorption of silicon at this energy, only the fluency of 300 μJ/cm² gave a good signal-to-noise ratio. The overall temporal resolution was about 50 fs at 1 kHz repetition rate. Around the transition wavelength of 810 nm (1.53 eV) between the visible and the IR region, the spectra are strongly affected by a strong signal arising from the residual light used for white light generation. A similar disturbance is observed in the spectra around the pump wavelengths.

The change in the absorbance, ΔA, may be modulated by light trapping effects particularly in the visible and IR ranges [16]. For this reason, we do not compare absolute intensities of ΔA across different wavelengths. Nonetheless, the dynamics of these signals still reflect the carrier dynamics of bulk silicon as these effects only influence the amount of energy deposited in the sample and not the dynamics.

## 3. Results

Figure 2 presents an overview of the results obtained with $E_p$=3.06 eV and 4.5 eV. They are depicted in the form of a 2D false color map, where the change of absorbance ($\Delta A = -\log(I_{ex}/I_0)$, where $I_{ex}$ and $I_0$ are the intensities of the transmitted probe with or without pump excitation ) is depicted as a function of probe energy ($E_{pr}$, x axis) and the time delay between pump and probe pulses ($\Delta t$, y axis). $\Delta t$ is reported on a logarithmic scale to highlight the short time range. Figures 2(a) and 2(b) show the results in the visible-UV region for $E_p$=3.06 and 4.51 eV, respectively, while Figures 2(c) and 2(d) show the result in the IR spectral range. As the pump is strongly scattered by the NW array, at $E_p$=3.06 eV, the residual pump signal is presented as a black bar. A negative value for ΔA corresponds to an increase of the transmitted light hence to an absorbance bleaching. Consistent features are observed when the wires were pumped at 3.06 eV or 4.51 eV with a slight difference in intensity. Three main spectral regions are recognized, 1.1–1.8 eV, 1.8–3 eV, and 3–3.5 eV. From 1.1 to 1.8eV, the signal increases (ΔA positive, yellow-red) in the first 1 ps after excitation, followed by a long decrease period. In contrast, from 1.8–3eV, ΔA decreases (ΔA negative, purple-blue) at early delay times after excitation, showing one minima of broad structures at 2.5 eV, and increases (yellow-red) at long pump-probe delays. In the UV range, a negative signal is observed in the whole time range, particularly intense in the 1-10 ps region.

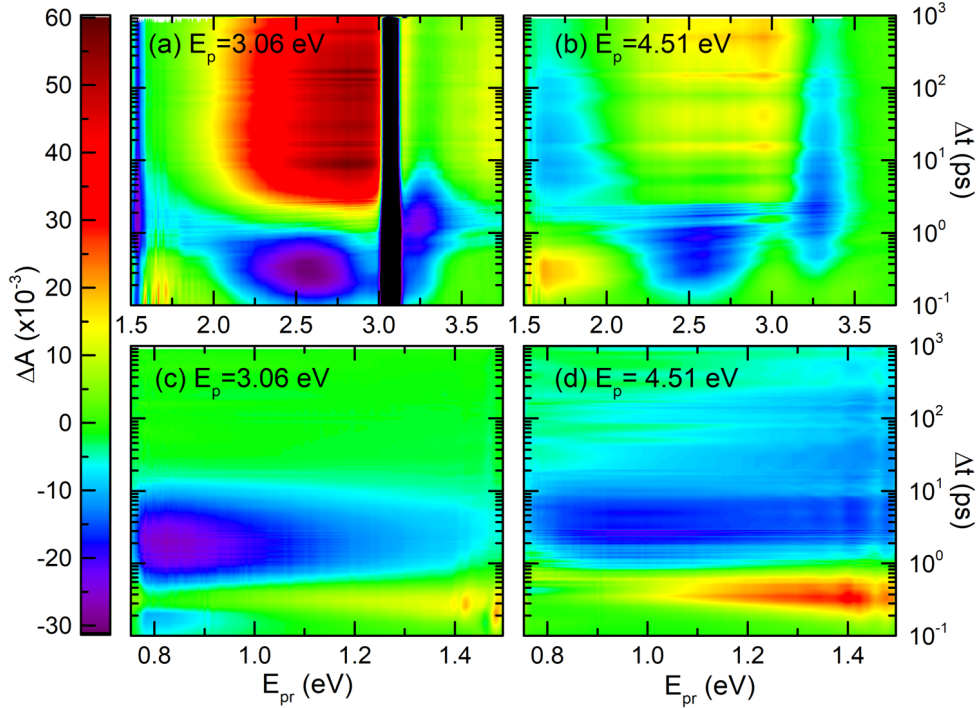

**Figure 2.** Two-dimensional false color map of the absorbance difference ΔA of Si NWs as a function of probe energy ($E_{pr}$, x-axis) and of the delay time between pump and probe pulses (Δt, the y-axis is on a logarithmic scale). (a) Pump energy, $E_p$=3.06 eV, visible/UV region; (b) $E_p$=4.51 eV, visible/UV region; (c) $E_p$=3.06 eV, IR region; (d) $E_p$=4.51 eV, IR region.

A more detailed spectral dependence of ΔA in the vis/UV region is reported in Figure 3 at different pump-probe delay times. We show the results for both pump energies, 3.06 eV (Figure 3(a) and 3(c)) and 4.5 eV (Figure 3(b) and 3(d)). The broad features around 2.5 eV reaches a negative maximum intensity much faster around 0.3–0.55 ps and change sign after 1–2 ps depending on the probe energy, similarly to what is observed in the transient reflectivity in the visible region in high quality thin film Si layers [17]. Then, the positive signals persist for longer than 900 ps. The narrow dip at 3.3 eV remains instead always negative with gradually reducing amplitude. This narrow absorption bleaching is observed at the energy corresponding to the direct optical transitions in Si ($E_1$, see later). Figure 3 also shows an important feature, namely the monotonic blue shift of the ΔA minimum around 3.3 eV with increasing delay as detailed in Figure 4. The amplitude of the blue shift is 133±15 meV for $E_p$=4.51 eV, and 72±12 meV for $E_p$=3.06 eV. 80% of the shifts occur in the first 2–3 ps. The dip energy reaches its maximum value around 3.32 eV at long delay times.

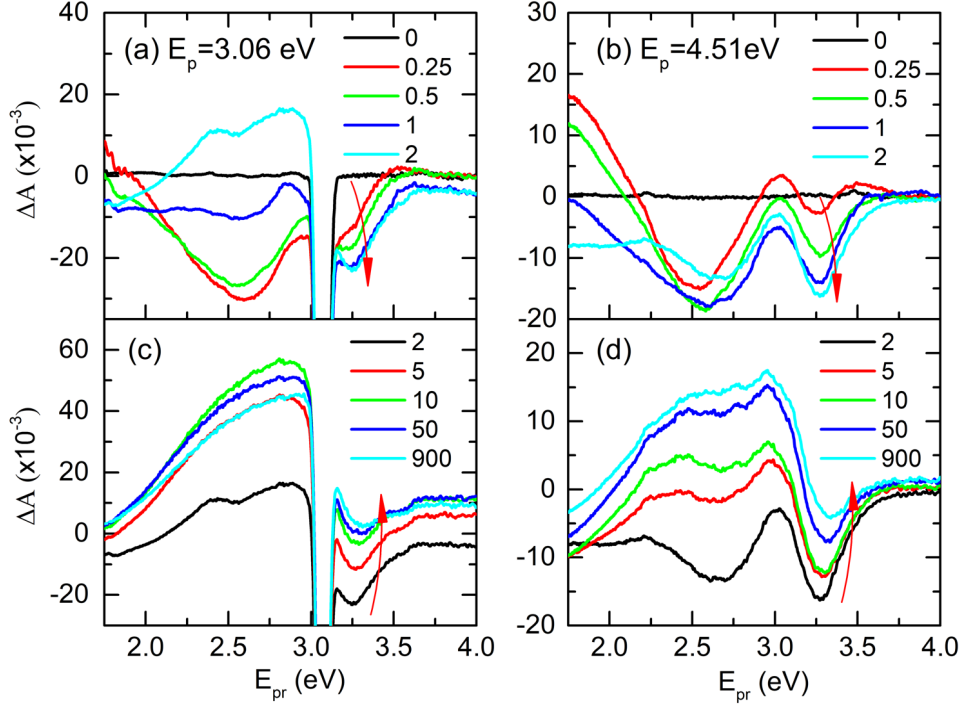

**Figure 3**. ΔA spectral dependence of Si NWs for different pump-probe delay times (Δt, in unit of ps) acquired with optical pumps of 3.06 eV [(a) and (c)] and 4.51 eV [(b) and (d)]. The red arrows indicate the direction of increasing delay times.

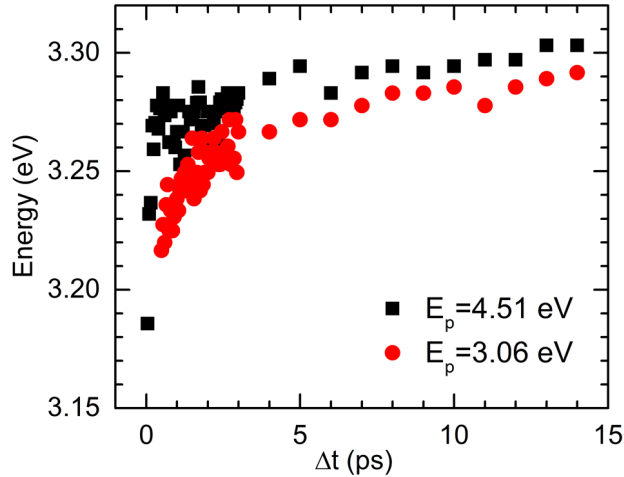

**Figure 4**. Position of the ΔA minimum relative to the direct band gap as a function of the pump-probe delay time for $E_p$=3.06 and 4.51 eV.

With the aim of disentangling the electrons dynamics in the conduction band above the $\Gamma_{15}$ minimum from that of the holes in valence band, we have performed the FTAS measurements also using a pump wavelength of 530 nm, corresponding to 2.34 eV, 1 eV less than the value of the direct band gap (3.3 eV). The corresponding 2D false color map is shown in Figure 5(a). The temporal evolution of the FTAS spectrum in the visible region is reported in Figure 5(b) and 5(c). Despite the pump energy being lower by 1 eV than that relative to the direct band gap, a clear bleaching of the absorbance is observed at the corresponding energy of 3.3 eV with a weaker signal than those obtained with the two other pump energies.

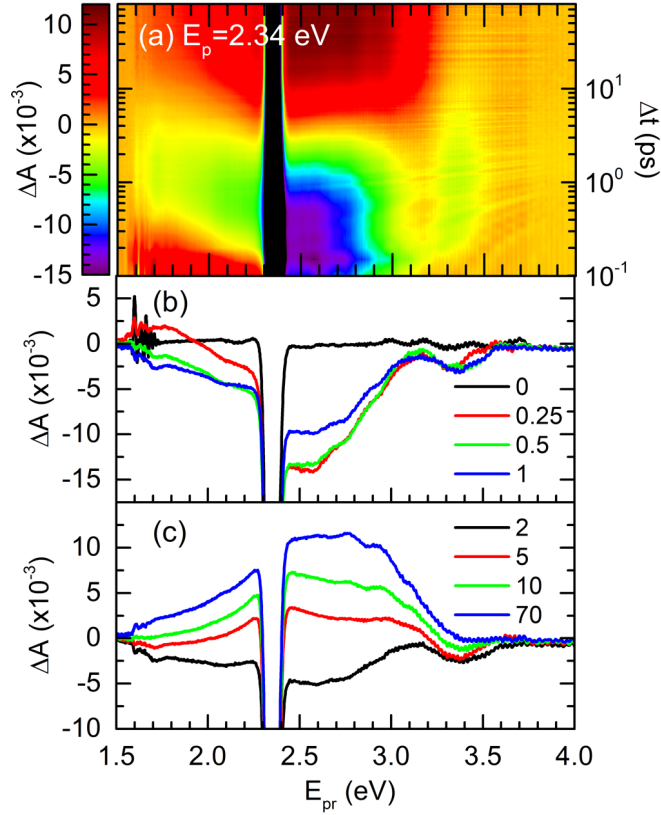

**Figure 5**. Results for $E_p$=2.34 eV. (a) Two-dimensional false color map of the absorbance difference, $\Delta A$, of Si NWs as a function of probe energy ($E_{pr}$, x-axis) and of the delay time between pump and probe ($\Delta t$, the y-axis is on a logarithmic scale). (b) and (c) $\Delta A$ spectral dependence of Si NWs for different pump-probe delay times ($\Delta t$, in unit of ps).

As an example of the temporal behavior of the $\Delta A$ signal, in Figure 6 we report the normalized values of $\Delta A$ at $E_{pr}$=3.3 eV for all the three pump energies used.

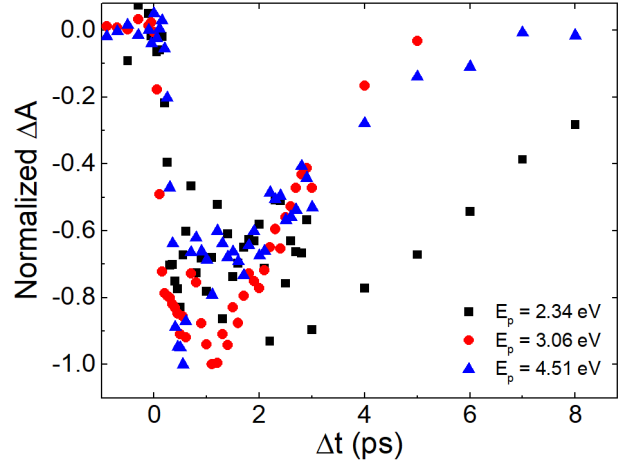

**Figure** 6. Temporal behavior of ΔA as a function of the delay time at $E_{pr}$=3.3 eV for all the pump energies used in the experiment. Values are normalized.

Finally, in Figure 7, we report the decay time measured at different probe energies for all the three pump energies used in this work: 4.51 eV (Figure 7(a)), 3.06 eV (Figure 7(b)) and 2.34 eV (Figure 7(c)). In the latter case, because we are mainly interested in disentangling the electron and hole dynamics in clearly determined bands, we only show the values taken for probe energies around the 3.3 eV signal.

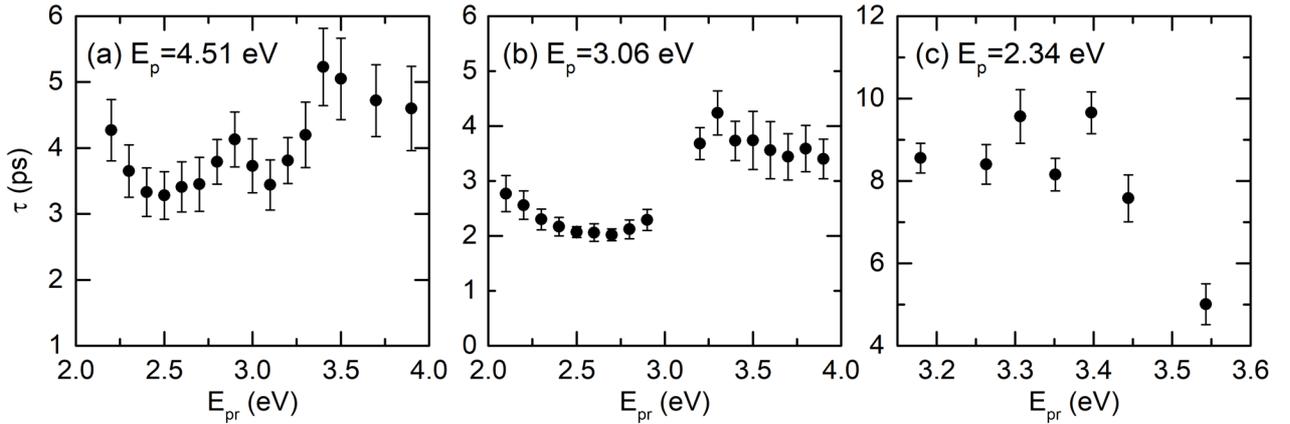

**Figure 7**. Decay time at different probe energy above 2 eV when the Si NWs are pumped at (a) 4.51 eV; (b) 3.06 eV and (c) 2.34 eV.

## 4. Discussion

### 4.1. General aspects

The absorption coefficient of silicon [18] shows a very smooth increase above the indirect band gap, remaining below $10^5$ cm$^{-1}$ up until 3 eV. For higher energies it sharply increases in correspondence of the $E_1$ and $E_2$ critical points (named after Aspnes and Studna[18]) of the band structure at 3.3 and 4.2 eV, respectively, corresponding to direct optical gaps at the Γ point of the Brillouin zone. In our experiments we have used three pump energies, 2.34 eV, 3.06 eV and 4.51 eV, the first energy being about 1 eV smaller than the direct band gap, the second one being smaller but close to the $E_1$ point, while the third one is larger than both the optical gaps corresponding to the $E_1$ and $E_2$. If it has been suggested that, due to the thermal broadening of absorption at room temperature and the strong photonic field generated by the pump, $E_p$=3.1 eV can also excite carriers at the $E_1$ critical point at Γ[5], this is certainly not possible when the sample is excited by the pump at 2.34 eV. The third pump, at 4.51 eV, will instead excite electrons from the valence band to both direct band gaps, above $E_1$ and $E_2$. The probe energy, instead, spans the entire energy range from below the indirect band gap to above the energy of $E_1$. The shape of the absorption coefficient is reflected into the pump-probe TA spectra that show broad features above the indirect band gap (1.12 eV) and a sharp dip around 3.3 eV (see Figure 2), corresponding to $E_1$. For energies below $E_1$, free carriers are generated along the Γ–X (above 1.12 eV) and Γ–L (above 2 eV) lines of the Brillouin zone. These free carriers are generated via indirect transitions involving phonon generation or absorption, with the creation of free holes at the top of the valence band at the Γ point and free electrons along the Γ–X and Γ–L line. These are the same indirect transitions that will be probed when $E_{pr}$<$E_1$.

After excitation by the pump electrons and holes very quickly lose phase coherence with the pump pulse and reach a quasi-equilibrium state in a thermalized Fermi-Dirac distribution with the characteristic temperature generally higher than the lattice temperature, also referred as to "hot carriers".

*4.2. The absorption bleaching at 3.3 eV.*

We first discuss the results relative to the direct band-gap at 3.3 eV. With the injection of free carriers, the corresponding absorption at occupied states is reduced, namely absorption bleaching is observed as negative ΔA signals. For all pump energies, a narrow dip at about 3.3 eV, corresponding to $E_1$, is clearly distinguished and lasts the whole time range (900 ps). This negative signal originates from Pauli blocking, as discussed in the cited literature [1, 4-9]. As we observe the bleached absorption also with $E_p$=2.34 eV, we can conclude that in this case it is mainly due to the pump-induced depletion of the electron population (or the creation of holes) at the top of the valence band around Γ. There is hence no need to create electrons in the conduction band to observe absorption bleaching involving transitions related to that conduction band.

A first feature observed in the absorption bleaching at 3.3 eV is its energy blue-shift with increasing delay time between pump and probe for $E_p$= 3.06 and 4.51 eV. This blue shift can be explained as the reduction of the band-gap renormalization (BGR) [19] with time due to the decrease of the density of the photoexcited carriers with the ensuing reopening of the band gap. Our results hence show that also BGR contributes to the spectral behavior of carrier dynamics. The insurgence of the band-gap renormalization occurs at pump-probe delays too short to be observed with our temporal resolution (about 50 fs). The Coulomb interaction between carriers is subject to screening especially when free carrier density is above the critical density ($N_{cr}$=6x10$^{18}$ cm$^{-3}$ for Si [5]). The large density causes overlapping of carrier wave functions, strong screening and thus decreased band-gap energy, leading to downward (upward) shift of the conduction (valence) band edge. The narrowed band gap recovers when carrier density reduces. As Figure 4 shows, with increasing pump-probe delay time, the energy of the ΔA dip gradually increases till it reaches a plateau after tens of ps, reflecting the one-electron $E_1$ value.

As mentioned above, because of our temporal resolution (about 50 fs) we are not able to observe the red shift of the direct band gap expected right after the optical excitation. This is in agreement with the result reported by Schultze and coworkers [7] who extracted a value of 450 attoseconds as the upper limit for the carrier induced (indirect) band-gap reduction from the temporal shift of the $L_{2,3}$ absorption edge. This is a much shorter time than that observed in other materials. In ZnSe NWs, as an example, we measured 170 fs as the rise time constant of the band-gap red shift due to band gap renormalization [20].

Using a widely used relationship between light intensity and photogenerated carrier density in films of solar cells [21] we can estimate the carrier density generated in the nanowires with the different pumps. With a pump fluence of 300 μJ/cm$^2$, approximately 6.96x10$^{19}$ cm$^{-3}$ e-h pairs are generated with $E_p$=3.06eV, and 9.12x10$^{20}$ cm$^{-3}$ with $E_p$=4.51 eV, respectively:

$$N_0 = (1 - R) I_0 \alpha \eta / h\upsilon \qquad (1)$$

where $I_0$ is the excitation density (J/cm$^2$), R is the reflectance, α is the optical absorption coefficient (1.22x10$^5$ cm$^{-1}$ for $E_p$=3.06 eV and 2.32x10$^6$ cm$^{-1}$ for $E_p$=4.51 eV, respectively [18]), $\eta$ is the generation efficiency, assumed to be unity, i.e. one photon generates one and only one e-h pair, and hν is the pump photon energy. R is deduced from reflectivity measurements (not shown) and is found to be much lower than in bulk Si [16].

In principle, by using the estimated photoexcited carrier density, the measured red shift of the bleaching signal due to this density, and the known hole effective mass, it is possible to obtain the effective electron mass of the $\Gamma_{15}$ conduction band minimum, which has been never estimated in experimental work nor explicitly published in theoretical papers. As described by Hartmann and coworkers [22], the bandgap narrowing is determined by carrier density, N, and carrier temperature, $T_c$. They gave a simplified and manageable relationship:

$$\Delta E_g(N, T_c) = -\frac{4.64(Na_B^3)^{\frac{1}{2}}R_{exc}}{\left[Na_B^3 + \left(\frac{0.107k_BT_c}{R_{exc}}\right)^2\right]^{\frac{1}{4}}} \qquad (2) \qquad (2)$$

Where $a_B$ is the excitonic Bohr radius, $R_{exc}$ is the excitonic Rydberg energy, with $a_B = \frac{m_0\varepsilon_r}{\mu}a_H$, $R_{exc} = \frac{\mu}{m_0\varepsilon_r^2}R_H$, where $\varepsilon_r$ is the relative permittivity of Si, $a_H$ is the Bohr radius of the hydrogen atom, $R_H$ is the Rydberg constant, and μ is the excitonic reduced mass, $\mu = \frac{m_e m_h}{m_e + m_h}$, with $m_e$ and $m_h$ being the effective masses of electrons and heavy holes, respectively.

Assuming $T_c$=1000 K as upper limit, we have for $E_p$=3.06 eV: $\Delta E_g$=72 meV and N=6.96x10$^{19}$ cm$^{-3}$. From these values, we estimate μ=0.052 $m_0$ and $R_{exc}$=3.86 meV. For $E_p$=4.51 eV we have $\Delta E_g$=133 meV and N=6.03x10$^{20}$ cm$^{-3}$ and we find μ=0.045 $m_0$ and $R_{exc}$=3.38 meV. If we use $T_c$=300 K, we find μ=0.051 $m_0$ and $R_{exc}$=3.80 for $E_p$=3.06 eV and μ=0.045 $m_0$ and $R_{exc}$=3.37 meV in the case of $E_p$=4.51 eV.

Therefore, a variation of $T_c$ by 700 K causes negligible change to the estimated results. We point out that the values obtained for the two pump energies are very close, with a difference of less than 10%, although they have been deduced by red shifts that were experimentally measured independently of each other.

Using the average value of those given above and using the heavy-hole effective mass of Si ($m_{hh}$=0.49$m_0$-0.537$m_0$) [23-26] we obtain $m_e$≈0.052 $m_0$ as the effective mass of the $\Gamma_{15}$ electron. No theoretical value for the $\Gamma_{15}$ electron mass has been published. However, the effective mass estimated in the described way is about 5 times smaller than the mass derived from the data used to calculate the Si band structure shown in Figure 1(a) of Ref. 5. From those data, indeed, we obtain 0.28$m_0$ for the electron mass as obtained by the second derivative of the relative density of states at the Γ point. The origin of the disagreement between the mass value obtained with the method described above and the theoretical value is likely due to the strong assumption made in the above calculations about the carrier density that is optically generated by the pump, obtained setting $\eta$ =1 in formula (1). To obtain the same shift of the band gap with an electron mass five times larger than that estimated, as suggested by the theoretical work cited above, we must have a carrier density about five times smaller (i.e. ~1.4x10$^{19}$ cm$^{-3}$ e-h pairs with $E_p$=3.06eV, and ~2x10$^{20}$ cm$^{-3}$ with $E_p$=4.51 eV) ), i.e., it should be $\eta$≈0.2. Formula (1), used to estimate the carrier density, is commonly used to design photovoltaic devices [21]. Hence, care must be taken when using it to design NW-based solar cells.

As the absorption coefficient at 530 nm (2.34 eV) is 10 times smaller [18] than at 405 nm (and about 200 smaller than at 275 nm(4.51 eV)), while the reflectivity is higher [16], a lower density of photoexcited e-h pairs is created in the NWs. The resulting value of 3x10$^{17}$ cm$^{-3}$ is lower than the critical density (see above) and a negligible BGR is expected in this case, where it would only involve the valence band. Indeed no shift is observed of the absorption bleaching in the time span typical of one-electron band gap recovery, see figures 5(a) and 5(b).).

Finally, it is worth noting that Hartmann formula relies on a universal behavior, i.e. the band gap renormalization is a function of the ratio of the exciton volume with the volume occupied by an electron-hole pair in an e-h plasma. This approach, yet simple and powerful, neglect details in the electronic structure that could affect the quantitative agreement between theory and experiments.

*4.3. Energies between 2 and 3.3 eV.*

Below the direct band gap energy, for short time delays, we observe a broad negative signal for energies above 2 eV. At short delay times, ΔA increases in negative intensity for increasing energy and shows a relative minimum around 2.5 eV (Figure 2). ΔA remains negative in the first 3 ps after excitation, the lower the probe energy the shorted the time span of the negative signal, while it becomes positive after that delay, indicating the occurrence of photo-induced absorption. The dynamics of ΔA in that energy range is much faster than that observed for the direct gap around 3.3 eV. In the literature a broad signal above 2 eV has been reported for both Si bulk [1, 9, 27] and nanowires [8]. In those works, that signal was attributed to hot-carrier relaxation with an important contribution from hot electrons along the Γ–X and Γ–L lines. However, on the grounds of the previous discussion of the absorption coefficient and our experimental results, we suggest that the absorption bleaching is mainly due to the fact that electrons at the top of the valence band are no longer available for excitation to states along the Γ–L and Γ–X lines of the Brillouin zone, similarly to what is discussed above with regard to the experiment with the pump at 2.34 eV and probe energy above this value. The broad bleaching signal reaches maximum amplitude in about 0.35-0.45 ps. This is a similar timescale to that predicted for electrons to relax from the Γ point to the L and X valley minima (~ 600 fs) in bulk Si [9, 28]. However, we note here that phonon- creation or absorption is also involved in probing any indirect transitions from the top of the valence band to the conduction band. The observed rise time is therefore also compatible with the picture where, after excitation, electrons at the top of the valence band are no longer available for excitation to states along the Γ–L and Γ–X lines.

For $E_{pr}$>2 eV, ΔA becomes positive at around 2 ps for $E_p$=3.06 eV, 3 ps for $E_p$=4.51 eV, while it takes longer (~4 ps) for $E_p$=2.34 eV., indicating photo-induced absorption. This sign change may be attributed to the re-excitation of carriers along Γ–L or Γ–X lines to higher energy states [24, 28] but this behavior requires further investigation for a more detailed explanation.

At energies below but very close to the direct band gap bleaching the signal becomes positive already at very short times (0.25 ps, see figure 3). This could be due to a carrier-induced deformation of the band structure that may increase the absorption at the energy close to the band gap minimum and creates a sort of positive wing in the signal lineshape.

A point that remains open is the reason why a relative minimum is observed at 2.5 eV, an energy that cannot be easily explained using the Si band structure. We point out that our results compare well with those of the literature for bulk-Si (when such results exist), showing that the use of NWs with diameter of 80 nm (or larger) has no qualitative effect on the carrier dynamics.

*4.4. Energies between 1.1 and 1.8 eV.*

As it can be observed in Figure 2, below $E_{pr}\approx$1.8–2 eV (the precise value depends on $E_p$) $\Delta A$ shows a different behavior than for higher energies, being positive in the first ps and then becoming negative. A positive signal indicates an increase of the absorption due to the pump excitation. Such a photoinduced absorption may be related to the presence of defects in the material that become populated after the pump excitation and give rise to a secondary excitation by the probe [29]. This possible explanation is supported by the fact that a fast positive signal is also observed in the first ps even for probe energies below the indirect band gap of silicon ($E_{pr}$<1.1 eV), The late occurring of the negative bleaching at these low energies suggests a competitive behavior between bleaching and photoinduced absorption.

*4.5. Relaxation times.*

As noticed above, the diameter of the NWs investigated in this work (about 80 nm) is large enough to not show an increase of the surface recombination rate due to thin diameters [8,10-13]. For pump energies close to the optical gap or above, we observe that the lifetime of the absorption bleaching is of less than 1 ps for $E_{pr}$<1.5 eV and increases to 2–4 ps for $E_{pr}$>2 eV, see figures 7(a) and 7(b), with a faint dependence on energy. The values instead observed for $E_p$=2.34 eV and $E_{pr}$>3 eV are higher (8±2 ps, figure 7(c)). As this pump energy is well below the value of the optical band gap, we assume that the measured decay time is only relative to the hole population dynamics in the valence band, because the number of electrons excited by the pump above the $\Gamma_{15}$ conduction band minimum is negligible. On the other hand, the dynamics for $E_{pr}\geq$3 eV is faster in the case of both $E_p$=3.06 eV and $E_p$=4.51 eV than for $E_p$=2.34 eV. This difference could be partially due to a reduced carrier-carrier scattering and/or Auger recombination because of the much lower carrier density excited with $E_p$=2.34 eV. As shown in Figure 6 the decay time observed at 3.3 eV is very similar for the $E_p$=3.06 and 4.51 eV, differences are within the error. Similar results are obtained for $E_{pr}\geq$3.2 eV. As the photoexcited carriers density is higher for Ep=4.51 eV than for Ep=3.06 eV, this feature suggests that the measured value is the characteristic, limit time of the many body effects involved, as the mentioned carrier-carrier scattering and Auger process.

As the decay time of the bleaching will reflect the fastest carrier dynamics between the electrons and the holes, we infer that the electron dynamics around and above the $\Gamma_{15}$ conduction band minimum is faster than the hole dynamics, as expected because of the mass difference, and hence it determines the temporal behavior of the absorption bleaching when both electrons and holes are photogenerated by the pump above the direct band gap minimum.

As mentioned above, we have observed that the dynamics with $E_p$=3.06 eV is very similar to that observed with $E_p$=4.51 eV and faster than that seen with $E_p$=2.34 eV. This feature suggests that although smaller than the direct band gap, the energy $E_p$=3.06 eV is sufficient to populate the $\Gamma_{15}$ conduction band, probably because of the contribution of both thermal energy and strong optical field, as suggested in reference 9.

## 5. Conclusion

In summary, we have performed femtosecond transient absorbance measurements on Si NWs grown on transparent substrates in the 1.1–3.5 eV range, spanning from the indirect to the first direct band-gap of silicon. Our work reports data that were partially missing not only in the literature regarding Si NWs but also in the overall literature on carrier dynamics in silicon. Beyond the description of the carrier dynamics over the whole energy range from the indirect to the direct band gap, we have shown the temporal behavior of the direct band-gap. We have shown the role of the band gap renormalization in the carrier dynamics and, using a pump energy well below the direct band-gap value, we have disentangled the hole contribution to the relaxation time of the absorption bleaching relative to direct band gap $E_1$. This disentanglement has also shown that the electron dynamics appear faster than that of the holes. Moreover, the possibility to investigate the whole energy range between the indirect and direct band-gap has allowed us to describe and discuss a more detailed picture for carrier dynamics of Si. Finally, we have also pointed out how the evaluation of the electron effective mass at the $\Gamma_{15}$ point is critically affected by the estimate of the photoexcited carrier density and that the evaluation of the carrier density in Si NW films is less trivial than the way it is usually done in thin films with a typical relationship linking light intensity, absorption coefficient and material volume. The presence of a clear minimum around 2.5 eV remains instead an unexplained feature. We hope that our experimental results, which represent a comprehensive study of the carrier dynamics in Si, could stimulate some new theoretical investigation.

**Acknowledgments.** This work has received funding from the European Union's 7th Framework Programme for research, technological development, and demonstration under Grant No. 316751 (NanoEmbrace) and from the Horizon 2020 program of the European Union for research and innovation, under grant agreement no. 722176 (INDEED). We are deeply indebted to Davide Sangalli and Andrea Marini (ISM-CNR) for having provided us with the data that form Figure 1(a) of reference 5 and for very fruitful discussions.

This is the version of the article before peer review or editing, as submitted by an author to Nanotechnology. IOP Publishing Ltd is not responsible for any errors or omissions in this version of the manuscript or any version derived from it. The Version of Record is available online at https://doi.org/10.1088/1361-6528/ab044a